\documentclass[12pt]{article}
\usepackage{graphics}
\usepackage{amsmath}
\def\={\phantom{..} = \phantom{..}}

\def\+{\phantom{..} + \phantom{..}}
\def\>{\phantom{..} > \phantom{..}}
\def\<{\phantom{..} < \phantom{..}}
\def\-{\phantom{..} - \phantom{..}}

\def\Sch{Schr{\" o}dinger}
\def\Schists{Schr{\" o}dingerists}
\def\Schism{Schr{\" o}dingerism}
\def\Schist{Schr{\" o}dingerist}
\def\Poin{Poincar{\'e}}

\title{\bf \Sch\ Was Right!\\[1in]
 He left his ``wavefunction" program incomplete,\\
 but nonlinear math and chaos theory can finish the job\\[3in]}

\author{W. David Wick}

\begin{document}
\maketitle
\pagebreak
\footnote{email: wdavid.wick@gmail.com}
\pagebreak

\section{Introduction\label{introsection}}

Now that we have reached the centennial of Erwin \Sch's seminal paper introducing
the wavefunction theory of matter, it is right and proper to inquire as to its legacy.
It is undeniable that today every paper in atomic physics cites his 1926 equation in the first paragraph.
But the philosophy undergirding the wavefunction seems to have fallen into the shadows.

This essay will take the form of a polemic combined with revisionist history. (I
mean revisionist in interpretations, not in questioning factual accounts.)
Thus I will pass swiftly through the key chapters in the dispute over the meaning
of his wavefunction, make some (sceptical or heretical) asides,
and finally arrive at the question: was his explanatory program a complete success?

No need to keep the reader in suspense: I will answer, not in \Sch's lifetime.
There were paradoxes and lacunae left in the paradigm, some of which \Sch\ himself described.
But I will argue here that recent developments in nonlinear mathematics, including so-called
``chaos theory", permit finishing the task. It turns out that one nonlinear addition to his equation from 1926
can resolve both the Measurement Problem and the Randomness Problem.
With this emendation, the wavefunction alone suffices to explain
the outcomes of many experiments (and it is particles that can be relegated to the shadows).

\section{1925-7: Revolution}

It all started with a paper in 1925 by a German physics student named Werner Heisenberg, who claimed
to have a theory of atoms that did not refer to any unobservable quantities---a command
central to the once-popular positivist philosophy, which declared certain hypotheses {\em verboten}
as theorist's fantasies.\footnote{Including, at one time, atoms!} Heisenberg displayed a set of equations describing an atom that
could only exist in one of a discrete list of energy states (i.e., states you could list like floors 
in an apartment building: the ground state, the next-higher energy state, and so on).
As would be made clear two years later in another manuscript (Heisenberg's famous ``Uncertainty Principle"
manifesto), he believed his equations described particles that jump from state to state,
or even from place to place.\footnote{Heisenberg included two illustrations:
one showing a continuous curve, which he said
was for planetary orbits, and a second in which the curve was replaced
by a series of dots, for the atomic realm. Jumps between discrete energy states were
proposed originally by Bohr in 1913.} Hence the label: ``quantum mechanics" for this thesis,
the first word signifying discreteness or countability.

Then, a year later, Erwin \Sch, a Viennese mathematical physicist, produced a radically different
theory, based on a seemingly antithetical view of atoms: 
that they are not a collection of particles 
waiting to jump, but rather are like vibrating drum-heads or guitar strings.
That is, oscillations or wave motions, the mathematics of which were already familiar (developed
in the previous century into a discipline called Fourier Analysis). It was known
that in such systems a discrete list of oscillations could 
always be found: the ``fundamental tone", 
the ``first overtone", and so on. These were mathematical constructions; no claim
was made that drum-heads can only vibrate in a single ``pure" tone, let alone
jump from one such state to another. The Fourier method could describe an arbitrary motion
by a combination or ``superposition" of those on the list.

Heisenberg had postulated his equations as the simplest he could imagine
susceptible to solution on the writing pad. But in another paper
in 1926, \Sch\ {\em derived} the same equations from his wave theory. That word is not
insignificant in the mathematically-based sciences. Indeed, in a plausible counterfactual,
after 1926 Heisenberg admits that \Sch\ lapped him, drops the jumping-particle business,
and joins the wavefunction program. (Of course, that didn't happen, and was perhaps a
psychological impossibility.) Instead, in the 1927 interpretative paper he brought
in a magic word from another field (psychology, insurance, or perhaps sports betting):
``uncertainty".\footnote{Heisenberg also used ``indeterminism" and other such terms in his paper;
even assuming
accuracy in the translation, discovering his precise meaning is very hard.}
His celebrated Uncertainty Principle served to excuse his inability to reproduce,
in a jumping-particles model, a simple fact
known from Fourier Analysis: that to make a wave-packet narrow in space, you have to combine a
lot of waves with different frequencies, and vice versa. In Heisenberg's mind
it became something about one's inability to know several aspects of a particle,
such as its position and momentum, simultaneously.\footnote{Or they don't possess
these properties? No one today really knows what Heisenberg was getting at.
In my first book, in 1995, I included four interpretations of what
Heisenberg had meant by his Uncertainty Principle. Reviewers of the book all claimed I
 didn't understand the UP, and produced their own versions. Needless to say, they didn't agree.}

Copenhagen (home to Niels Bohr, Heisenberg's mentor) thought the wavefunction an abomination.
They had to have their jumping particles. \Sch\ would not capitulate, reportedly exclaiming,
when pressured by Bohr, that if he had known ``this damned quantum jumping" was going to persist,
he would never have gotten involved in the atomic business in the first place.

Late in 1926, the Copenhagenists settled on their interpretation of the wavefunction (in opposition
to \Sch's belief that it is an element of reality, like Einstein's gravity tensor). 
A collaborator of
Heisenberg, Max Born, interrupted calculating the wavefunction for a scattering problem
to proclaim: ``... {\em only one interpretation is possible} ...  [the wavefunction]
gives the probability
[of where the electron goes]." (My italics.)
But in a footnote he remarks: ``...more careful consideration shows the probability
is proportional to the square... ." Neither makes sense; Born was overlooking that
the wavefunction has ``complex values", mathematical jargon for a pair of real numbers dubbed
``real and imaginary parts".\footnote{Archaic terminology from the 19th Century.
Better would be ``two-dimensional numbers",
avoiding any hint of mysticism.} Meaning that it cannot be a probability, which is a single, positive real number.
Nor can its square, which might be negative. Copenhagen soon cured that disease, declaring that
the modulus squared (same as length, squared) of the wavefunction at each point represents ``the probability
of the particle being there".\footnote{After 1935, when Einstein and two colleagues,
Podolsky and Rosen, attacked the Copenhagenist doctrine
that Quantum Mechanics gave a complete description of a particle, it became ``...where the particle
would be found, or found to be going (if observed somehow)".}

I take a moment here to assess the logic, or meaning, of Born's declaration. First, the assertion
that some interpretation is the only one possible is belied by the history of science.
Interpretations fluctuate with the times; ``atom", ``element", ``species" all have different meanings now
than in centuries past.\footnote{And, as we will see, ``particle"!} Second, I think that Born's statement can be further elucidated as:
``... only one interpretation is possible {\em that permits continuing to believe that I am
learning something about a particle, even while I am computing a wavefunction...}"
Thus invoking probability let Born absorb a foreign notion into his system,
as some polytheistic religions incorporated other people's gods.
And third, what did he mean by ``probability"? There is no agreement about what the words
``chance", ``likelihood", and ``probability" are supposed to mean.
Did Born refer to an objective measure, such as averages computed from repeated trials? If so,
why don't all the trials produce the same outcome, in which case
the resort to probabilities would be unnecessary?
Or did he mean the subjective interpretation,\footnote{This is the most common
interpretation of ``probability". If I say, ``The probability that the Seahawks
(a football team resident in Seattle) wins tonight's game is 0.7,"
I am not referring to a long series
of identical match-ups in identical circumstances, from which I have computed the win-rate,
but rather to my current opinion. So no one can gainsay it, even if the Seahawks lose.}
 as representing ``my opinion (or knowledge, or ignorance,...)
about where the particle is headed?" If so, when did ``I" become so important?

\section{The War is On}

So the two camps parted company. One became known as Copenhagenism.
The other I will call \Schism, and it is the creed which I espouse
(although \Sch\ might not have approved every detail of my version).
But before continuing the story,  I must dispose of some Copenhagenist doctrines advanced
by Bohr which lead only to utter confusion.
I refer to ``wave-particle duality" and ``complementarity".

These conceptions officially entered into physics in the winter of 1927, following a generational spat.
After a skiing vacation, Bohr returned to Copenhagen
and discovered that his younger colleague had just sent a manuscript to a journal.
Bohr then dropped an Alpine boulder on his prot{\'e}g{\'e}. Heisenberg recalled in a
memoir published many years later that ``it ended with my breaking out
in tears because I just couldn't stand the pressure from Bohr."\footnote{From [PRICECHISSICK].
In my cinematic imagination, Heisenberg stands in front of Bohr with his fingers
crossed behind his back and mutters, {\em soto voce}, ``They jump!”}
He forced Heisenberg to add an Addendum to his paper; here is part of what he had to write:

\begin{quote}
Bohr had brought to my attention that I have overlooked essential points in the course
of the discussion in this paper. Above all, the uncertainty in our observation does not
arise exclusively from the occurrence of discontinuities but is tied directly to the demand
that we ascribe equal validity to the quite different experiments which show up in the
corpuscular theory on the one hand and in the wave theory on the other hand.
\end{quote}

Physicists, Bohr claimed, had been forced by the very phenomena they studied to
accept two conflicting pictures of the microworld. Neither can supplant the other;
you just have to live with it. He dubbed this new conception “the Complementarity Principle”.

What if scientists in past centuries had adopted this way of thinking?
In the 18th Century, taxonomists, confronted with a specimen that both
flew and suckled its young, did not invoke ``Aves-Mammalian duality" (``it's a bird in the air,
but a mammal on the ground!") to evade classifying.
Today one could imagine a Linnaeus or Buffon of physics creating a new family or order
 (Wavicles?) as the historical figures did for bats. But Bohr didn't want any new categories;
he was happy about the mystery.\footnote{There is convincing evidence from
memoirs of his colleagues that Bohr had
a philosophical agenda that long predated quantum theory; e.g., the ``four men in a boat"
story from Heisenberg. See [TIB], paperback edition (1996), p. 187.}

As for the neologism ``complementarity", in the 1960s the Irish physicist John Bell would
label it a euphemism, maintaining that what Bohr really meant was ``contradictarity":
the worshiping of paradoxes or contradictions absent any desire to resolve them.
Although I do enjoy reading the fascinating conceits of philosophers (Bohr thought he was one),
as guides for scientists' thinking, these notions are toxic. Also, they played no role
in the most important later developments, in my opinion.

There were other factions at that time (Einstein, endlessly searching for his ``light-quantum",
was a party of one), and over the succeeding decades many more were founded.
But amidst all that noisy paradigm-production, the core issues were elicited by two men:
Johnny von Neuman and Erwin \Sch.

\section{Cats All the Way Up}

First came von Neumann, a mathematician from Hungary who wrote a book about the mathematical
foundations of quantum mechanics in 1932. Taking the Copenhagenist interpretation
as writ, he codified the doctrine in a set of axioms.\footnote{Mathematicians need axioms
in order to contribute to science. But they often fail to realize
that somebody's axioms may simply embody the  doctrines of  one faction.
I noticed this tendency also among mathematicians interested in biology,
who usually adopt R. A. Fisher's model of population genetics from his book in 1930,
which neglected
predation and parasitism. Axioms, and models, should be taken with a large helping of salt.}
Included was one which asserted that an observation must yield a result on a
particular list of real numbers.\footnote{An astonishing proposal for anyone who
 has ever worked in the laboratory, or with real data.}
With admirable rigor he then examined exactly how such observations could be made, according to
his axioms. There he ran into trouble.

The problem was a little word, `or'. Suppose something can happen in one of two mutually-exclusive ways;
say, a pointer on a dial can move up, or it can move down.
Von Neumann's problem was that he didn't get the `or' in that sentence,
but rather `and'. Using \Sch's wavefunction description, he found two wavepackets:
one representing ``the pointer moved up" and another representing ``it moved down".
On the yellow pad, formulas for both wavepackets
appear added together, and erasing one is impermissible. What to do? Perhaps add an observer:
another device or even a person; but now you have wavepackets representing ``pointer went up,
device or person records that it went up" and the same, with down replacing up.
No progress towards an actual result.

Von Neumann despaired of finding anything in his axioms which could remove this embarrassment.
So, notoriously, he invoked ``the consciousness of the observer to collapse the wavefunction"
and yield a definite outcome.\footnote{I cannot read German, so I read the English translation
of the book made by Princeton U Press
in the 1950s, ``with the cooperation of the author". There I found not ``consciousness"
but ``Ego state"
of the observer.  How did Freud get into a book with the title:
{\em On The Mathematical Foundations of Quantum Mechanics?}}

Next came \Sch, in 1935, finally writing his interpretative essay.\footnote{Translated and
reproduced in [REDBOOK].} He wanted to make a point
similar to von Neumann's, but, being a literary man, he naturally created a metaphor or parable
(which became a cultural clich{\' e}).   But it was Rube Goldberg complicated and had one bit that
obscured the message.

He asked the reader to contemplate the fate of a cat locked in a steel box together with
a “diabolical contraption:” a Geiger counter equipped with a small piece of uranium and connected to 
a hammer and a flask of hydrocyanic acid. If a uranium atom disintegrates, the counter clicks,
the hammer shatters the flask, and the cat dies. Or it didn't and the cat lives.

The problem with this imagery is the iron box, which inevitably prompts people to contemplate
``my knowledge, or lack thereof” of the cat's fate. (``I get it! Until I open the box,
I don't know if it's alive or dead!" So? Many things are concealed from us, signifying nothing.)
Rather, \Sch\ was discussing whether the necessity of the wavefunction description
means that, in the microscopic world of atoms, things are “blurred” or “washed out”. Such ``blurring", he
demonstrated, could jump instantly to the macroscopic realm of cats and people.
He should have made the box out of glass, so we can see whether the cat's mortal state is really smeared out.

Another point of confusion is whether \Sch\ was announcing an exciting new discovery:
cats can be both alive and dead! \Sch\ was not; he was describing what philosophers of science
usually call, in polite language,  an ``anomaly" in the theory's predictions. But ``absurdity" is more like it.

Curiously, today the word ``cat" appears frequently in technical articles in quantum/wavefunction physics,
but the modern version is much simplified. Gone are the box, the hammer, and the flask of poison.
Rather, we just have a cat sitting beside a Geiger counter, whose sound level has been turned up.
When the counter emits a loud ``click", it startles the cat, who leaps, say, one meter to the left.
Call it ``\Sch's leaping cat", and anything resembling it is ``a Cat".\footnote{The
technical definition is: a ``Cat" is a macroscopic object whose dispersion of the center-of-mass,
computed from the wavefunction, is larger than the object itself.}

Von Neumann's vexation can now be recast as: when trying to use
this theory to understand observations,
it's Cats all the way up.

\section{What the Wavefunction Got Right}

I interrupt the theme of incompleteness to remark on two
striking predictions of \Sch's theory. I refer to interference
The term ``interference" entered into physics in the early 19th Century. The initial observation,
by British polymath Thomas Young, was of dark and light bands appearing on a card
when light was passed through two openings. This celebrated ``two-slit experiment"
could only be explained if light were a wave phenomenon,\footnote{If a wave crest
meets a wave trough at the screen, you get the dark band; if two crests or two troughs meet,
a light band. Two bullets can dig a deeper hole than one, but cannot cancel each other.}   and so
delivered the fatal blow to Newton's corpuscular theory of light. Then came James Clerk Maxwell
with his electromagnetic field postulate, then Louis de Broglie noting that electrons have wave properties,
and finally \Sch\ extending the scope of waves to all kinds of matter.

In 1927, British physicist G. P. Thomson sent ``cathode rays" (we now say electron beams)
through a metal foil
and saw the same kind of circular bands that you would expect if you sent light
through an optical grating.\footnote{Thomson received the Nobel Prize in Physics in 1937
for his electron diffraction experiments.
Curiously, G. P. was the son of J. J. Thomson, the recipient of the Nobel in 1906 for having shown,
back in 1897, that cathode rays were a stream of subatomic particles---the first to be discovered,
soon renamed ``electrons". Thus the father proved the electron was a particle and
the son proved it was a wave.}
Nevill Mott then calculated the exact pattern from \Sch's wave theory.
Later, in the 1950s, \Sch\ reminisced:

\begin{quote}
And, I don't regret to say, we surely need those spherical waves as realities
(not merely expressing our lack of knowledge), if we wish to account, e.g. for G. P. Thomsons's
beautiful experiments on the interference of de Broglie waves ...  one finds that no pattern
of thought has been discovered ... without regarding the wave functions ...
as describing something real.
\end{quote}

For Copenhagenists, interference remained a bug-a-boo. Richard Feynman maintained
that the two-slit experiment was ``the only mystery in quantum physics". If asked, which
slit did the particle pass through? he would refuse to answer.\footnote{Feynman
was also a positivist, and so could justify his refusal to answer by pointing to the lack of
a ``particle detector" situated at the slit.} \Schists, of course, would reply,
``Particle? {\em What particle?}"

``Entanglement" as a possibility originated in a
remark \Sch\ made at the end of his cat piece. Perhaps his most
prescient observation, it is now considered crucial for building the much-hoped-for
``quantum computer". The issue concerns how we describe the parts of a whole.
Consider, for example, a chair consisting a wooden back and legs, a leather seat, and metal screws.
Each of the parts are described by their physical characteristics: their dimensions, chemical compositions,
states of motion, and so forth, as is the whole. That was the classical understanding of matter.
But, \Sch\ says, in his theory the parts of a composite object {\em do not have their
own wavefunctions}.
In modern phraseology, the parts are ``entangled" in the wavefunction of the whole.
This was a possibility never dreamt of in anyone's philosophy, prior to \Sch.

Ironically, interference and entanglement are sometimes described as the two most characteristic features
of ``quantum theory". But if anyone had really tried to make a theory of atomic phenomena
based on jumping particles,\footnote{Coincidentally, I happen to have expertise
in a branch of mathematics, created after the
1920s, called ``stochastic jump processes". I used them for some problems in biology.
The equations are completely different from those of Quantum Mechanics.}
 they would never have imagined either possibility.

\section{Many Voices Shouting}

The new ``quantum" or ``wavefunction" theories were clearly defective.
Copenhagenism made an unjustified call upon probability theory, and buried contradictions
under a heap of metaphysical detritus. Neither paradigm could explain observations.

Given this degree of cognitive dissonance, its not surprising that the many proposals
made since the 1920s range from the serious to the silly, from sane to utterly bonkers.\footnote{A
particularly striking moment occurred in the 1950s, when Wolfgang Pauli wrote in a letter
to Max Born (included in The Bohr-Einstein Letters)
that he thought perceiving an object in a definite position
must be ``regarded as being a `creation' outside the laws of nature"---in other words, a miracle!}
In 1995, I published a survey of this literature, and  am not tempted to repeat the task.
(Of course, it has grown immensely since then.)
It will suffice here to state my personal motivations and biases.

My goal has always been to drive metaphysics out of physics.\footnote{That was of course
the goal of the positivists, but they went too far, ruling out any element of
an explanation that they could not see with their own two eyes.} My mantra is: Stop philosophizing
and write better equations!

What must these ``better equations" accomplish? Well, what is the ultimate source of our vexation?
It is the Superposition Principle.

The SP, which justified von Neumann's adding together those wavepackets,
is a consequence of a most peculiar aspect of Heisenberg's and \Sch's equations:
their unrestricted linearity. Not in any other science with a basis in mathematics
is it claimed that a linear equation represents anything fundamental. Waves in the ocean can be
described by a linear equation; but it is an extremely crude approximation to what is going on with those
zillions of H$_2$O molecules. The equation even fails for water in a shallow channel.\footnote{In
narrow, shallow channels, water can exhibit strange wave forms called ``solitons"
which move without spreading out. They are described by nonlinear equations.}
The Cat problem---conventionally referred to as the Measurement Problem (MP),
because the state of the cat, leapt or non-leapt, can be thought of as measuring the state
of the uranium atom---derives entirely from the SP.

We must break the hold of the SP on physics. So my second mantra is: Nonlinear is necessary!

\section{\Schism\ Defined}

Although my first mantra demands one stop abusing philosophy, I won't pretend not to have one.
I now relate the
\vskip0.1in

\centerline{\bf Tenets of \Schism\ (Wick's Version)}
\vskip0.1in

1. Matter is described by a wavefunction.\footnote{The wavefunction is a complex-valued
function of 3N spatial variables plus time, with N the number
of component bodies constituting the system.}

2. The wavefunction has nothing to do with probabilities.
Nor is it some kind of statistical function.\footnote{Einstein believed that
the wavefunction represented an ensemble, presumably of
his ``light-quanta" or some similar thing representing electrons.}
Nor is it a measure of ``my ignorance" of a particle's whereabouts.
Nor does it possess entropy.\footnote{Von Neumann thought an individual wavefunction
possessed entropy, but strangely it would be
constant over time, violating the Second Law of Thermodynamics.}
It is an element of reality.\footnote{One cannot write ``it is real"
due to that archaic real-and-imaginary jargon for complex numbers.}

3. There are no ``quantum jumps" or ``quantum leaps". The wavefunction evolves continuously,
according to some nonlinear equation.

4. When discussing measurements, the wavefunction must be generalized to cover
the measuring device. Having done so, any further observers, including humans, are redundant.

5. To make contact with data, it is only necessary that
the macro (device) observables be defined as functionals\footnote{``Functional"
is mathematical jargon for a function of a function. For example: in elementary calculus,
the integral of f(x) over [a,b] is a functional of a function.} of the wavefunction.

 The last two tenets smack of another philosophy, called ``instrumentalism";
but they really just follow from experience. Data in the physical sciences are not
lying around waiting to be collected, like fungi on a forest floor;
they are created by apparatus. Which do not self-assemble in the night.
They are designed and built by the experimenter, as prompted by her paradigm.\footnote{Data,
as Kuhn told us in {\em Structure} (1962), are not theory-free. That remark deep-sixed
philosophies of science from Francis Bacon through the positivists.}

\section{Aside: ``Particles" and That Other Cat}

How can my tenets leave out ``particles"? Surely today's Standard Model that explains
everything represents the triumph of ``particle physics"?

I'll let you in on a secret, actually known to every knowledgeable person in the field
(excepting science journalists): ``particles" aren't what they used to be.

The beginning of the end of ``particles" as we knew them was, of course, 1926,
with \Sch's first paper on wavefunctions.
Appearing in every wavefunction are ``arguments": variables such as x, y, z,
representing coordinates in space. But  these do not describe a particle at this spot;
rather, the wavefunction has a value at every such
triple. The wavefunction is at every place at once; in fact, in many-body contexts,
it is {\em at every N-tuple of places at once}. In \Sch's world,
there simply are no ``particles".\footnote{At the start \Sch\ thought that a ``particle"
might be a very narrow wavepacket;
but he gave up the idea when he realized there was nothing in his theory
to keep it from spreading out.}

What was a particle supposed to be before 1926? It's essence lay in its localization. A wave may be
everywhere at once, but a particle is {\em over there}.
Born's probability hypothesis was basically an attempt
to maintain some kind of localization: he could not say where the particle is,
but could at least say
that, perhaps with probability 0.7, it is in the top half of the box.

If the reader is thinking, ``Abandon particles? That's a bridge too far for me,"
 I hasten to point out
that \Schists\ are not alone in their abnegation. According to conventional histories,
beginning in the 1930s  there was a second shake-up in physics.
This episode, called ``the second-quantization revolution", involved making
creation and annihilation
of ``quanta" in ``fields" the central elements. But American Professor Steven Weinberg
(Nobelist in 1979 for helping to complete the Standard Model)
maintained the phrase was a misnomer. The episode was really about finally
killing off that despised wavefunction.\footnote{Weinberg remarks, in Volume I of his
three-volume treatise entitled {\em The Quantum Theory of Fields},
that physicists at the time did not believe that a wavefunction theory of finitely-many particles
could satisfy the Relativity Principle.
But a Russian named Vladimir Fock showed later it could be done
with an indefinite number of particles.}  But he admitted that, by taking an ax to
\Sch's big idea,
they also cut out any notion of localized particles. Thus the
Standard Model might accurately describe what comes out of the collision chambers at CERN,
but only in the sense of ``the meson emerged heading towards Geneva with energy so-and-so"
superposed with ``it headed towards Pouilly with energy such-and-such".

What is left for the poor particles? Here is what a Nobelist said:\footnote{Weinberg, 1996.}

\begin{quote}
``... particles are just bundles of energy and momentum of the fields."
\end{quote}
So ``particles" in the literal sense disappeared from physics---like that other
celebrated fictional cat,
only the grin remained.

\section{The Randomness Problem, Part I}

We have met the Cat Problem, better known as the Measurement Problem, and know
why it demonstrated
that \Sch's program was incomplete. Another such indication derives from the apparently random
outcomes in many experiments in atomic physics. Sometimes ``particle detectors"
chime like bells in a cathedral tower whose ringers are tossing dice. Call it
the Randomness Problem (RP).

We have seen how the Copenhagenists finessed the difficulty by calling on probability,
but never acknowledged the ambiguity involved. So, what interpretation of probability
should physicists adopt?  Certainly not the subjective; why should my opinion matter
to what transpires in an atom? The proper concept for use by scientists, I believe, was given
by French mathematician and physicist Henri \Poin\ in a book published in 1896.
It is sometimes called ``the dynamical interpretation".  \Poin\
asked why we resort to probability in certain situations, and not in others.
He traced its use to instabilities in some parts of a whole,
plus the inability to precisely control some variable.\footnote{Keynes, in his 1920 tome,
disparaged \Poin's work as of
``no theoretical importance". How can elucidating the physical origin of a concept
be of no importance?
Keynes own theory of probability is obscure and justly forgotten.}

Consider roulette. Ignoring that scene in {\em Casablanca} where Rick lets Captain Renault win,
we hope the croupier cannot control the toss well enough to produce a predetermined outcome.
The job of the spinning wheel is then, besides confounding the croupier's will, to blow up
and simultaneously smooth out randomness (e.g., so all outcomes are equally likely).

\Poin\ is famous for his many contributions to mathematics and physics.\footnote{Among
other things, \Poin\ invented the mathematical subject called ``topology" and beat
Einstein by a year to one of his famous formulas. (Not {\em that} one,
but it was only one step away.
Einstein was lucky to get the credit.)}
Relevant here is that he almost discovered so-called
``chaos theory" by pure mathematical contemplation,
unaided by an electronic computer.\footnote{The insight stemmed from a mistake in a manuscript
submitted to a Swedish Academy offering
a prize. \Poin\ won with a piece about the three-body problem in astronomy.
The prize was a publication;
but a Swedish academician wrote him questioning one assumption.
\Poin, horrified, replied, in effect: ``Stop the presses!" The issue concerned instabilities.}
I'll describe shortly how chaos can resolve the RP.

\section{Blocking Cats, Part I}

In 2017, this author proposed that the MP could be solved by a fairly simple modification of
\Sch's equation from 1926. The idea was that the new terms would be
essentially inactive on the microscale of atoms or nuclei, where we know that
ordinary linear theory works well. But on the macroscale the modification would nullify
the Superposition Principle, which really differentiates the two worlds. And, of course, block Cats.

How could that be accomplished? First, we need an aside about energy.

\section{Aside: What is Physics?}

If a high-school student contemplating taking an advanced-placement course in physics
asked me, ``What is physics? What do physicists study?" I would answer with one word:
energy.

Modern physics began with Newton in the 17th Century. His postulated attraction between
distant bodies represented a new form of energy (although the concept had not yet jelled).
Thermodynamicists in the middle 19th Century reinterpreted heat in a body,
previously thought to be a sort of fluid, as due to the energy of constituent molecules in motion
(despite  the atomic theory of matter being still controversial). The great Scottish mathematical physicist,
James Clerk Maxwell,  postulated that electromagnetic waves transported energy
By the end of the century, energy---appearing in multitudinous forms,
passing from one to another, but at no point created or destroyed---had
become established as a universal concept.

In his second 1905 relativity paper, Einstein proclaimed the victory of energy over the last hold-out,
writing: ``the mass of a body is a measure of its energy content".\footnote{A certain
well-known equation first appeared in this work.}
Thus the energy concept took over everything. Any new theory in physics will inevitably
involve modifying or articulating the energy.

\section{Blocking Cats, Part II}

Besides the stated goal, I drew two lines in the sand (or on the yellow pad) that I would not cross.
One was energy conservation: I would not entertain a theory that did not predict that energy is conserved,
in every possible history of the system, exactly. This of course reflects my personal prejudice
about what is physics and what is not.

My second derives from the famous Einstein quote: ``God does not throw dice." My take on what he meant
by this aphorism is, in physics jargon, ``A fundamental theory cannot
contain any stochastic elements." By ``fundamental" is meant underlying, providing the ultimate explanation.
``Stochastic" refers to random choices that are made over time, like
throwing dice to chose the next move in a board game. Einstein was reminding his colleagues
that, if probability methods are used, it must be because the underlying dynamics is
too complicated to represent exactly. A stochastic theory can only be some kind of crude approximation
to a fundamental theory.

Having fixed my biases, I next needed a mechanism that would attain the goal.
The simplest I could imagine is called ``an energy barrier". This scenario is very familiar to physicists
and engineers: for example, the reason that traveling
to space is so expensive is the cost of supplying the energy necessary to climb out of Earth's
gravity well. So I modified \Sch's equation to create an energy cost
to making a Cat (remember: a beast that is pulled apart
over a length greater than its size; i.e., duplicated, like the leapt-and-didn't-leap cat).

Introducing a new form of energy requires defining its strength through a
parameter, which I denoted by the letter `w'.\footnote{I chose `w' with no malice aforethought.
When writing a long math paper, you usually
run out of letters, partly because many are already assigned; e.g., `c' is the velocity of light,
`e' is the charge on the electron, etc. Most of the Greek letters have standard meaning, too;
e.g., `psi' denotes the wavefunction. The only one left as I wrote the introduction
to that paper was `w'.}
As an illustration, I invented a suitable magnitude for `w' and calculated the leaping-cat energy,
which came out so great the creature nearly attains the speed of light.
If that energy were converted into heat, the poor cat would morph into a glowing plasma.
Since no such energies exist in terrestrial laboratories, Cat formation is impossible.
Meanwhile, for a few atoms or nuclei, this energy is literally infinitesimal and can be ignored,
explaining why ordinary linear theory works so well on that scale.

How did the theory attain these goals? By incorporating dispersion into energy and
exploiting ``non-standard scaling".  ``Scaling" is engineering jargon for how quantities
change when you compare equivalent systems of different sizes. Physicists usually assume
that energy scales proportionally to the size of the system, which we might indicate
(as if we are particle-ists) by ``N", the ``number of particles".
So energy is supposed to grow proportionally to N. The trick with the new form of energy is that,
in the presence of a dispersed object, it scales as N squared.
Since N for a macrosystem is huge---say,
around Avogadro's number---that factor would be stupendous, requiring 46 digits
before the decimal point to write it out. Say goodbye to Cats.

This proposal was admittedly novel. In the history of physics, energy
mostly had an interpretation in terms of the relative positions of objects or of their velocities.
Everyone recalls from school Newton's law of gravitational force
acting mutually on two ponderable bodies: proportional to the product of their masses,
and inversely proportional to the square of the distance between them. Coulomb's law for electrostatic
force is similar, except that charges replace masses and, if alike, repulsion replaces attraction.
Both laws imply a form of energy called “potential”, because when bodies move in response to the force,
an invisible form of energy is converted into observable “kinetic energy”.
Quantum mechanics took over these energy notions intact.\footnote{Einstein,
in a letter to Max Born, disparaged the quantum theory as an attempt
to describe the world in terms of rubber bands regulated by potential energy.}

But my new form of energy has no interpretation in terms of particles;\footnote{See
[WICK-LAGRANGIAN].}
rather, it represents an aspect of the wavefunction as a whole.
If (to coin a phrase) ``Heisenbergist energy"
is about the spread of particles or how quick they are (to jump I suppose),
``\Schist\ (non-conventional) energy" is about the spread of a wavefunction or
how quickly it can expand or contract.\footnote{There are two forms for WFE
I considered: one based on dispersion in space, and another based
on dispersion of momentum. Which is best suited to the job is still under investigation,
as I write, in January 2026.} So, from now on, I will refer to the novel addition
as ``wavefunction energy", abbreviated WFE.

\section{The Randomness Problem, Part II}

After proposing a solution to the MP by introducing new terms into \Sch's equation,
I turned to the RP. Both \Sch's original theory and my modified version are competely
deterministic (no dice-throwing), so my first thought
was to assume wavefunctions have a universal random element. I then checked that this
kind of dynamics could manifest a violation of Bell's Theorem, a famous result from 1964 that distinguishes
the predictions of quantum mechanics from classical mechanics, even
permitting the latter some unknown, hidden variables.\footnote{The exact statement is that,
if the predictions of quantum mechanics
are true, then local hidden variables are ruled out by Bell's Theorem. My theory with a random
wavefunction component can violate it, because the ``hidden variable" is the wavefunction itself,
and it isn't a localized object.}  These results constituted the
second paper in my \Schist\ program.  Then I remembered so-called ``chaos theory",
which was a hot topic in mathematics 40 years earlier.

What I recalled from attending all those seminars back then was that high-dimensional,
nonlinear dynamical systems are often ``chaotic". So, in a third communication,
I abandoned the ``universal random component" idea.

What is ``chaos"? From its etymology it once meant ``a state of total disorder"
(as for, e.g., the universe the moment  before God said, ``Let there be ...").
But after a best-selling book appeared in the 1980s,\footnote{See [GLEICK].}
it acquired a completely-different meaning, referring to a new branch of mathematical physics.
This field, which developed in the 1960s, treats of the predictions of certain dynamical theories.
There are various definitions, but ``chaos" primarily seems to entail three things:
(a) sensitive dependence on initial conditions (``the butterfly flapping its
wings today in Seattle causes a hurricane in Brazil a century later"); (b) a qualitative distinction
between behavior on different time-scales, and (c) random-appearing trajectories.
The easiest example for illustrative purposes comes from
computational astronomy.

 Thirty years ago, a group simulated the motions of the outer planets of
the Solar System on a fast computer and compared
two trajectories, the second obtained by making a 1.5 cm shift in the center-of-mass of Uranus.
The trajectories diverged on a time-scale of millions of years and, if sampled
(say by space aliens repeatedly flying past the SS) the planets' positions
would appear to be somewhat random. While on a human time-scale, say of centuries,
the motion of these planets is completely predictable and regular.

In a series of papers I studied the chaos question, first for simplified models
of a few ``qubits" (the type of systems currently being developed to serve as the basis of
``quantum computing"). Using standard criteria,\footnote{I showed that,
 when a certain matrix criterion was satisfied, the ``Lyaponuv exponent" was positive,
meaning that nearby trajectories diverge exponentially.} I showed that a 3-qubit model with WFE
was indeed chaotic, in sense (a) and (b) above. By contrast, conventional linear theory
generates ``Lissajoux figures": combinations of simple sine curves with different frequencies.
Such figures may look exotic to the non-mathematician, but are not chaotic.
I also showed how a ``criterion for chaos"\footnote{It is not known if the criterion implies
a positive Lyaponuv exponent.} can be extended to models set in the continuum.

Thus, experiments with atoms may resemble trips to the rouelette table:
outcomes vary not at the whim of {\em Tyche},\footnote{Greek goddess of chance;
Roman: {\em Fortuna}.}
 but because we cannot completely exert control
at the bottom level.  (Usually only a few components of the wavefunction
are under the experimenter's thumb.)

\section{Satisfying Einstein}

Even if Einstein could have overlooked the incongruence of the wavefunction
with his ``light-quantum" from 1905, \Sch's theory of 1926 would not have pleased him.
For it was not compatible with his Relativity Principle (also dating from 1905).

In the vernacular, the RP says that no observer, by virtue of state-of-motion,
has a privileged right to state the laws of physics. Thus physicists living on
a moon of Jupiter should believe the same laws as Earthly physicists.
More technically, it demands that an equation take the same form when written in variables taken
from any ``reference frame", whether stationary or moving. (\Sch's wasn't.)

Two years later, the English mathematician Paul Dirac wrote another wavefunction equation,
which was.\footnote{Dirac's wavefunction had four components. This expansion was chosen
in order to incorporate
an electron's ``spin" and to solve the compatibility problem.} This development initiated the march down the long and winding road that lead eventually
to the Standard Model. (Along the way, wavefunctions, including Dirac's, were abandoned
by most of the community.)

My modified \Sch's equation with WFE was also not compatible with Einstein's 1905 Principle.
But in that first paper I showed that a Dirac-style version with a momentum form of WFE was
``relativistically-invariant", as physicists write. To motivate a momentum version,
recall the \Sch\ leaping cat: it both leaps and doesn't leap. After a while, it
becomes a cat that sits a meter to the left and a cat that sits where it was.
Thus, in order to create a spatial leaping Cat,  it may be necessary first to create a momentum Cat. 
In which case, blocking the latter also blocks the former.

In 1915 Einstein introduced his theory of gravity, which replaced Newton's force acting
between distant bodies by ``space-time curvature". As one step, he generalized his Relativity Principle
from referring only to special coordinate systems to an arbitrary such system;
the former then became ``Special Relativity" and the latter, ``General Relativity".
In a later paper, I  generalized my alternate WFE to this broader regime of curved space-times
(with some limitations), and to have ``General-Relativistic invariance".\footnote{See [WICK-GR].}

I hate to brag, but the Standard Model has never been shown to be compatible with GR.
Nor has anyone constructed a so-called ``quantum gravity"
theory which would replace Einstein's with something making a Copenhagenist happy.\footnote{If
the reader Google's ``quantum gravity", she will get a zillion hits. But be skeptical;
I claim none of these attempts can derive any interesting consequence of Einstein's theory, e.g.,
Black Holes. Indeed, most of them cannot even predict the sign
(whether gravity is attractive or repulsive).}

\section{Strange Statistics}

Another triumph of the wavefunction derived from counting cases, assigning probabilities, and
excluding possibilities. Nobel Prizes resulted.

Recall that unfortunate rule you learned in your college probability or statistics class
about counting equivalent cases. The one that says: if you cannot distinguish the cases,
they must all have the same probability.  Despite its ubiquity, this is probably the worst
of all definitions of the term.\footnote{In his big tome from 1920,
John Maynard Keynes (more famous as an economist)
disparaged it as “The Principle of Indifference”  and enumerated the ways this rule
can get you in trouble.
(Keynes was hoping to do for probability what Russell and Whitehead were trying to do for logic,
with pretty much the same outcome.)} Let's see what it says about some simple situations.
Suppose you flip two coins. How many cases? Most of us would say, four: HH, HT, TH, and TT.
Then, according to the rule, the probability of a match is 2/4 = 1/2.
But why isn't it three cases: both heads; one head and one tail; and both tails?
Then the probability of a match is 2/3.

Why do we go with the first choice? Presumably
because we believe that coins are distinguishable. Perhaps one has a slight chip on an edge, say.
So we go with four cases: H(smooth)H(chipped), H(chipped)T(smooth), H(smooth)T(chipped),
T(smooth)T(chipped).

But now let us consider waves. Suppose you are watching ocean waves arriving at a rocky shore.
Sometimes you see one crest and spray, sometimes two, and sometimes none.
Would anyone count the second event, for probability purposes, as two cases: crest A and crest B,
or the other way around? Wave crests, if they are of the same size, we regard as indistinguishable.
So only three cases.

So the key is ``distinguishability". This issue arose in atomic physics when electrons
were attributed a characteristic called ``spin". Wolfang Pauli, and then Paul Dirac,
produced wavefunction theories with more components which represented ``spin states". These
were assumed to take just two values, call them ``up" and ``down".
Now the counting-cases business arises.

But we aren't talking about coins but rather wavefunctions!
Recall that a wavefunction is a complex number associated with arrangements of things.
If you interchange some of these things, two cases arise:  it could stay the same, or it could be
replaced by its negative.\footnote{There are other logical possibilities,
like being multiplied by `i', the imaginary unit,
but for technical reasons they aren't allowed in quantum physics.}
  In the first case, you can have as many identical things (“particles”)
in a given state as you wish.  In the second, the wavefunction must be zero, because the sign
changes if you switch them, and even for complex numbers only zero can equal its negative.
So indistinguishable things can't occupy the same state. The wavefunction of
several  electrons obeys the second rule; so it must vanish on states:
``up,up" and ``down,down".  Thus, wavefunction physics predicts that (indistinguishable)
electrons never occupy the same state,
a conclusion which generated a Nobel for Wolfgang
Pauli (it's called the ``Pauli Exclusion Principle").

Now let probability enter, either through Born's {\em ad hoc} rule, or by imagining ensembles of
wavefunctions.\footnote{Even if random outcomes are due solely to sensitive dependence
on initial conditions, as
in my current theory, a probability distribution will be placed on those conditions. As
\Poin\ noted, the particulars of this distribution may not matter.}
It is apparent that different statistics will arise depending on
whether ``particles" (arguments of the wavefunction, for us \Schists) are distinguishable or not,
and whether they have ``spin". The names
of various physicists who explored the cases got attached (``Fermi-Dirac statistics"
for spin-possessing electrons; ``Bose-Einstein statistics" for spinless particles).

For particles-only people, these strange statistical scenarios must have come as a great surprise.

\section{Aside: Clicks, Drops, Dots, and Lines}

But isn't \Schism\ afflicted by apparitions? What about the click of the particle-detector?
The line of droplets in the Wilson Cloud Chamber (or the curly-cues in those modern pictures
made at particle accelerators)? That Dot on the Screen?

Probably in the mind's eye it's that last image that people find most troubling.\footnote{Ignoring
the issue of whether anyone has ever seen a single dot on a piece of photographic film or
a copper plate. Normally, experiments in atomic physics incorporate a beam directed at the screen;
to manage to produce exactly one registration would require turning down the beam flux,
waiting perhaps a long time, and getting lucky. That Dot on the Screen is primarily
a rhetorical device or image.}
But for me, it was the Cloud Chamber.

I once was privileged to be shown a Wilson Cloud Chamber. Somebody in the physics lab\footnote{I
do not remember where. It might have been at the University of Chicago,
where I was an undergraduate physics major; at the University of Washington, where I
was briefly a physics graduate student; or at Princeton, where I was a physics postdoc.}
had jury-rigged one from a multi-gallon jug filled with some supersaturated vapor
and equipped with a bit of radioactive material, likely an alpha-emitter, in the base.
A line of droplets appeared and I thought, ``Wow! I just saw a particle!"
It was more than a decade later, after I started my re-examination of  \Sch's philosophy, that
I realized that what I saw that day was not a particle. It was a collection of visible, macroscopic
objects; the mind then literally connected the dots.

Suppose you look upwards and see a long, linear pattern in the clouds: a line in the sky.
What is the explanation for it? The simplest is that it is a jet contrail. Even if you didn't
see the plane; maybe it went by earlier, or was flying too high to spot.
But one day, while hiking in the Olympic Mountains near my home
in Seattle, I saw not just a line but a row of teeth: regularly spaced lines
at right angles to the long one. Had the passing plane fired rockets at
regular intervals? Or did a goddess threw her comb in the air? Dismissing such whimsy, some
complex, nonlinear, nonequilibrium hydrodynamics must have been at work.

Which raises the specter of Alternative Explanations.

Here is English physicist Nevill F. Mott in 1929, explaining
that, in considering the Wilson Cloud Chamber, you should not imagine a wave representing the
alpha-particle that leaks out of the nucleus:

\begin{quote}
The difficulty that we have in picturing how it is that a spherical wave
can produce a straight track
arises from our tendency to picture the wave as existing in ordinary three-dimensional space,
whereas we are really dealing with [wavefunctions] in the multispace formed by the co-ordinates
both of the alpha-particle and of every atom in the [cloud] chamber.
\end{quote}

Mott goes on to show how two gas atoms ``cannot be ionized unless they lie in a straight line
with the radioactive nucleus."\footnote{Ionized atoms can nucleate droplet formation.}
That explains the linear tracks but note also that ``particles"
appear as aspects of the whole set-up; they are epiphenomena.

For the Dot on the Screen, we picture a ``photon" striking there, and then cannot
understand how a light wave of width covering the entire screen can possibly produce
a single exposed grain. But it's always the same story; the wavefunction does not
represent ``a photon" but the whole: light-plus-screen.  I have another mantra to
chant about these cases:\footnote{The knowledgeable reader may detect a similarity of this mantra
to a certain saying
attributed to Richard Feynman.}

\begin{quote}
 Every experiment in wavefunction physics can be explained, it turns out, by saying:
 ``You remember the Wilson Cloud Chamber? It's the same thing."
\end{quote}

In a paper from 2025 I presented a small-scale model of ``particle detectors" and a wavefunction;
simulating it, the Dot appeared. I showed also that Born's probability law was compatible with
the data generated.

Then there are the Spectral Lines, which appear when the light intensities given off  or absorbed
by electrified gases are plotted versus frequency (color). Don't the SL demonstrate that atoms
live in discrete energy states, occasionally jumping from one to another
while emitting or absorbing one of Einstein's ``light-quanta"?\footnote{In another of his 1905
papers,  Einstein suggested that light was ``emitted or absorbed
by atoms as a whole, and moved without dividing". This Patent Office clerk
was trying to resurrect Newton's corpuscular theory of light! It's as crazy as if
a chemist tried to revive phlogiston.}  Textbooks all say so.

First, the SL aren't lines; that is a graphical illusion resulting from scrunching the horizontal
(frequency) axis. They are local maxima (hills).\footnote{The Copenhagenists cope with the
line-width issue by invoking the ``Time-Energy Uncertainty Principle",
which (a) is another call on that magic word from psychology, and (b) doesn't exist
in the mathematics.} Second,  Alternative Explanations!
In the 1950s, \Sch\ suggested, but did not demonstrate,
that these bumps in the spectrum represent ``resonances": frequencies preferentially
absorbed or radiated. Later, in the 1980s, Professor A. O. Barut and colleagues
constructed a theory of atoms interacting with light (as Maxwell first described it),
and displaying such resonances in the spectra.\footnote{See [WICK-BARUT] and references therein.}
Finally, Einstein spent nearly five decades trying to find
a theory in which his light-quanta would appear, but failed. The ``photon" of the Standard Model
is not his light-quanta, as it is unlocalized.\footnote{The ``photon number" is really
just an integer that appears in that theory,
just as one appears in the theories of all kinds of vibrating systems.}

The simplest explanation is rarely the only explanation.

\section{Aside: Missing Those Quantum Jumps}

I realize that television and movie screenwriters will now have to think up more plausible transitions.
But it can't be helped; the “classic” quantum jump seems to have
been refuted by an experiment!

In 2019, Zlatko Minev and an international group of collaborators published a paper
with the astonishing title: ``To catch and reverse a quantum jump mid-flight."
Copenhagenists imagined their beloved jumps were instantaneous, unpredictable, and unstoppable,
like a “flash crash” in the stock market. The experimenters engineered an artificial “atom”
with three energy levels, two which can undergo a back-and-forth transition,
and a third which monitors and controls the process. Here are their conclusions:

\begin{quote}
The experimental results demonstrate that the jump evolution when completed is continuous ...
and deterministic. Furthermore, ... we catch and reverse a quantum jump mid-flight ...
It is possible to acquire a signal that warns of the imminent occurrence of the jump ... the results ...
support a contrary view [to that of Bohr's] consistent with that of \Sch\ ...
\end{quote}

Minev remarked to an interviewer from the Canadian magazine New Scientist: ``In a sense,
these jumps aren't jumps." In ordinary usage, a ``jump" refers to an event that might be observed
from start to finish, as in a skater's performance at the ice show.
The skater's partner might even catch her,
mid-flight. That's why announcers never proclaim, ``She made a quantum leap!"

It's just another case of Alternative Explanations. It was always possible
that what appears to be a jump
was really just a continuous transition (and the illusion created by compression
of the time-axis in a figure).

Nearly three decades before the publication of this experiment,
while sitting in the audience during a talk
by the British quantum-physics expert Anthony Leggett,
I recall thinking that, if anyone ever managed
to watch a ``quantum jump" from beginning to end, it would finally do in Bohr's philosophy.
I put it down to idle dreaming.

\section{Where's Your Evidence?}

I have rivals, whose proposals I usually reject on the ideological grounds mentioned earlier.
But on one issue, we are all in the same boat: lack of experimental evidence.

In 2019, I wrote a brief expository paper about what type of experiments might
yield evidence weighing for or against my theory.\footnote{Which sounds evasive, but experiments
can neither ``prove" nor ``falsify" a paradigm, as Thomas Kuhn argued convincingly in 1962.}
By simulating from a few-qubit model I produced a graph, essentially
Cat-iness vs. external potential applied to a system, with a hockey-stick shape.
Surveying the literature, I found a suggested experiment with all the right features.
An Anglo-German team proposed to turn a graphene wafer---a disk of carbon a few atoms thick and
smaller in diameter than a pollen grain---into a Cat.\footnote{See Abdi, M., {\em et al.}}
 But I did not find a follow-up
on the results of the experiment.

My original thought about the ``Infamous Boundary"\footnote{A phrase my co-author,
William Faris, suggested for the title of our 1995 book. He found it in a paper of
John Bell. It might mean the dividing line between where classical physics rules and where
it gives way to wavefunction (or ``quantum", if you prefer) physics.}
was that it resided at the scale of
apparatus. But in a 2025 obituary in the New York Times I read about the work of
the Nobel-Prize-winning chemist, Dr. Martin Karplus.\footnote{New York Times, 19 January 2025,
p. A 28.} The headline referred to Karplus
as the man ``Who Made Computers a Chemist's Tool", and contained this passage:

\begin{quote}
The models used classical Newtonian physics to predict how multitudes of atoms
and molecules move during reactions, and they used quantum physics to describe
how chemical bonds are broken and formed
during these reactions. This type of analysis proved particularly useful
in understanding biological reactions involving enzymes,
the proteins that govern chemical responses in living organisms.
\end{quote}

Why, I wondered, could classical physics serve for modeling some parts of molecules,
while ``quantum physics" was necessary for other parts? Where was the dividing line?

Then I found an article by two philosophers of chemistry with the remarkable title:
 ``The Problem of Molecular Structure Just is the Measurement Problem."
Apparently many macromolecules of biological importance come in two forms,
one the mirror-image of the other. The MP enters into explaining why we only see one
or the other species and never a superposition of the two.

 This recent history of chemistry suggested to me that the IB
might reside in macromolecules.\footnote{See [WICK-MACRO].}
I therefore sketched some experiments which may or may not be do-able by
real chemists.\footnote{I never worked in the chemisty laboratory,
and in the physics laboratory only briefly,
in college.}

And that's the sad status of the field. We are all pining for an Eddington to lead an expedition,
or at least for a 43 seconds of arc we might explain.\footnote{By retrodicting
the 43 seconds of arc per century shift in the perihelion of Mercury, Einstein
showed that his gravity theory of 1915 was more than a mathematician's dream. Arthur Eddington
lead the 1919 eclipse expedition that produced some (actually, minimal) evidence for the theory,
which made Einstein famous as the new Newton.}

\section{Completing \Sch's Program}

  I have shown that incorporating WFE into \Sch's equation can explain why classical
physics works so well on the macroscale, while wavefunction physics dominates on the
microscale. It also provided an explanation for why outcomes in some experiments
appear to be random which does not rely on an {\em ad hoc} invocation of probability, or
some philosophical stance.

Why, I sometimes wonder, didn't \Sch\ himself make some proposal along the same lines?
In his defense, I note that, in 1926, no electronic computers were available.
And the chaos scenario was not formulated for another four decades. I needed both
to complete his program.

In a paragraph above, convention requires that I should have written
``quantum physics" in place of ``wavefunction physics". By now, the reader has grasped
that I avoid using the q-word.\footnote{A recent aversion, I admit; my first five papers
in this area all
included the q-word in the title.} Why? Because I have no idea what ``quanta" are supposed to be.
The word harkens back to a an obsolete description of the microworld as
populated by particles that jump.
From my (admittedly biased) perspective, the word ``quantum" should be retired.

A lot more mumbo-jumbo can be dispensed with, too. E.g., ``It's a wave if you don't look,
but a particle if you do!" And ``When the Geiger-counter clicks or doesn't, the cat in the box
becomes two cats, one dead in this universe but alive in a different one!"\footnote{The reference
is to the ``multiverse", an oxymoron that appeared in physics after
a 1957 thesis by Hugh Everett, and has since become strangely popular.} Also, it's time to
boot out ``uncertainty", ``consciousness",  and any other words imported from other disciplines,
from physics.

I think \Sch\ would have approved my modest suggestions. In a lecture in 1952 in Dublin,
he said:\footnote{See [GRIBBIN], p. 222.}

\begin{quote}
Nearly every result [the quantum theorist] pronounces is about the probability of this or that ...
with usually a great many alternatives. The idea that they may not be alternatives
but all really happen simultaneously seems lunatic to him ... all contours becoming blurred
and we ourselves rapidly becoming jellyfish ... Still, it would seem that, according to the quantum theorist,
nature is prevented from rapid jellification only by perceiving or observing it ... it is a strange decision

\end{quote}

\section{Epilogue}

The argument that began that January in 1926 is still ongoing as I write. Perhaps a dozen factions
uphold doctrines that seem outlandish, if not lunatic, to physicists upholding others.\footnote{It
reminds one of Kuhn's many-paradigms-competing, Babylon phase, in which believers
in one paradigm use terms or concepts that are incomprehensible to believers in another. Kuhn
thought a revolution came next and swept away all but one paradigm. But the history of
wavefunction physics suggests that alternative paradigms may survive, even if (temporarily)
suppressed.} Metaphysical
or mystical excuses for ignoring ``anomalies" still find a following. Why haven't physicists found an escape
route from this quagmire?

All we really needed to drive metaphysics out of modern physics was a better equation.

\section*{Bibliography}

Historical surveys and article collections:

[TIB] {\em The Infamous Boundary: Seven Decades of Heresy in Quantum Physics},
by W. David Wick, with a mathematical appendix by William Faris. Birkhäuser, 1995
(with a different subtitle) and Copernicus, 1996 (paperback, with an index).

[LUDWIG] {\em Wave Mechanics.} Ed. Gunther Ludwig. 1968. Contains translations of
\Sch's 1926 papers.

[GRIBBIN]  {\em Erwin \Sch\ and the Quantum Revolution,} by John Gribbin.
John Wiley and Sons, Hoboken, NJ. 2013.

[BECKER] {\em What is Real? The Unfinished Quest for the Meaning of Quantum Physics, }
by Adam Becker. Basic, 2018.

[REDBOOK] {\em Quantum Theory and Measurement,} J. A. Wheeler and W. H. Zurek, Eds.,
Princeton University Press, Princeton, NJ, 1983. The essential ``red book" reprints and translates
from the German many of the original papers.

 [BELL] {\em Speakable and Unspeakable in Quantum Mechanics,} by John Bell.
Princeton, 1987.

[PRICECHISSICK] {\em The Uncertainty Principle and Foundations of Quantum Mechanics:
a Fifty Years' Survey.} John Wiley \& Sons, New York, 1977.

[BORN-EINSTEIN] {\em The Born-Einstein Letters, 1916-1955.} First published in 1971; reissued in
2004 by Macmillan.

[GLEICK] {\em Chaos: Making a New Science}, by James Gleick.  Viking Books, 1987.

Wick's papers on the Measurement Problem, the Randomness Problem, and related issues:

(All are available on the physics pre-print server, arXiv, and can be downloaded without charge.)

[WICK-MPI] ``On Non-Linear Quantum Mechanics and the Measurement Problem  I: Blocking Cats."
(identifier 1710.03278), October 2017.

[WICK-MPII] ``On Non-Linear Quantum Mechanics and the Measurement Problem  II:
The Random Part of the Wavefunction." (identifier 1710.03800),  October 2017;

[WICK-MPIII] ``On Non-Linear Quantum Mechanics and the Measurement Problem  III:
\Poin\  Probability and ... Chaos?"  (identifier 1803.11236), April 2018;

[WICK-MPIV] ``On Non-Linear Quantum Mechanics and the Measurement Problem  IV:
Experimental Tests." (identifier 1908.02352), published August 2019.

[WICK-GR] ``On Nonlinear Quantum Mechanics, Space-Time Wavefunctions, and
Compatibility with General Relativity." (identifier 2008.08663) August 2020.

[WICK-CHAOSII] ``Chaos in a Nonlinear Wavefunction Model: An Alternative to Born's Probability
Hypothesis." (identifier 2502.02698) Feb. 2025.

[WICK-CHAOSIII] ``Islands of Instability of Nonlinear Wavefunction Models in the Continuum:
A Different Route to 'Chaos' ." (identifier 2512.09109) December 2025.

[WICK-DotS] ``That Dot on the Screen: also, what about Born? and other objections
to wavefunction physics." (identifier 2504.17808) April 2025.

[WICK-BARUT] ``Nonlinear, Nonlocal: Comparing A. O. Barut's Theory to Mine, with special
emphasis to That Dot on the Screen." (identifier 2505.13704) May 2025.

[WICK-MACRO] ``Can the Infamous Boundary Be Found in Macromolecules? Also, Von Neumann
vs. \Sch\ Ensembles and `Hund's Paradox' in Quantum Chemistry."
(identifier 2506.02227) June 2025.

[WICK-LAGRANGIAN] ``Locality, Micro- vs. Macro-, Particle Interpretations and All That:
A Lagrangian Approach to the Measurement Problem." (identifier 2509.07206) September 2025.

Other quoted or discussed books and articles:

{\em The Mathematical Foundations of Quantum Mechanics}, by John von Neumann.
English translation of the original German text (1932). Princeton University Press, 1955.

``On the Quantum Mechanics of Collisions”, by Max Born. Zeitschrift f{\" u}r Physik, 37: 863-67, 1926,
translated into English  in [REDBOOK].

{\em Calcul des Probabilit{\' e}}. By Henri \Poin. 1896. A neglected masterpiece.

 Mott, Nevill F., ``The Wave Mechanics of Alpha-Ray Tracks,"
Proceedings of the Royal Society of London, A126, 79 (1929), reprinted in [REDBOOK].

\Sch, E., ``Are There Quantum Jumps?" Part I, British Journal for the Philosophy of Science
 V. 3 No. 10, p. 109-23, and Part II, V. 3 No. 11, p. 233-242. (1952).

Weinberg, S. {\em The Quantum Theory of Fields.} Three volumes. Cambridge U. Press, 1995. Vol. One
has a lengthy historical account of the development of ``particle physics."

Weinberg, S. The quote in the section about particles is from a speech 
given at a conference entitled
``Historical and Philosophical Reflections on the Foundations of Quantum Field Theory,"
Boston University, March 1996;  I accessed a version entitled ``What is a Quantum Field Theory,
and What Did We Think It Is?", on arXiv, published 4 Febuary 1997.

Szpiro, George G. {\em \Poin's Prize: The Hundred-Year Quest to Solve One of
Math's Greatest Puzzles.}
Plume, 2007.

Murray, N. and Holman, M. ``The Origin of Chaos in the Outer Solar System."
Science 19 March 1999, 283: 1877-81.

Franklin, A. and Seifert, V. A. ``The Problem of Molecular Structure Just Is
the Measurement Problem".
Brit J Phil Science, 75(1) 2024.

Keynes, J. M.  {\em A Treatise on Probability}. Cambridge, 1920. My version reprinted by
Rough Draft Printers, 2008.

Kuhn, Thomas S. {\em The Structure of Scientific Revolutions.} 3rd Ed. U of Chicago Press, 1962, 1970.

Abdi, M., Degenfeld-Schonburg, P., Sameti, M., et al., ``Dissipative optomechanical preparation
of macroscopic quantum superposition states." arXiv, May 2016; formal publication,
Phys. Rev. Lett. 116: 233604.  (2016).

\end{document}